%% file: main.tex
\documentclass[10pt]{sosp2015}

\usepackage{amsmath}
\usepackage{amssymb}
\usepackage{caption}
\usepackage{color}
\usepackage{graphicx}
\usepackage{hyperref}
\usepackage{listings}
\usepackage{microtype}
\usepackage{subcaption}
\usepackage{times}

\title{\coz{}: Finding Code that Counts with Causal Profiling}

\authorinfo{Charlie Curtsinger\titlenote{
  This work was initiated and partially conducted while Charlie Curtsinger was a PhD student at the University of Massachusetts Amherst.}}{
  Department of Computer Science \\
  Grinnell College}{
  \texttt{curtsinger@cs.grinnell.edu}}

\authorinfo{Emery D. Berger}{
  College of Information and Computer Sciences \\
  University of Massachusetts Amherst}{
  \texttt{emery@cs.umass.edu}}

\exclusivelicense
\conferenceinfo{SOSP'15}{October 4--7, 2015, Monterey, CA}
\copyrightyear{2015} 
\copyrightdata{978-1-4503-3834-9/15/10} 
\doi{2815400.2815409}

\microtypesetup{
  final,
  tracking=true,
  kerning=true,
  spacing=true,
  factor=1100,
  stretch=20,
  shrink=20
}
\SetTracking{encoding={*}, shape=sc}{0} 

\definecolor{mygreen}{rgb}{0,0.6,0}
\definecolor{mygray}{rgb}{0.5,0.5,0.5}
\definecolor{mymauve}{rgb}{0.58,0,0.82}
\lstdefinelanguage{c++threads}[]{c++}{morekeywords={pthread_create,pthread_join,thread,join}}
\newcommand{\coz}{\textsc{Coz}}
\newcommand{\punt}[1]{}

\begin{document}

  \maketitle
  
  \input{abstract.tex}

  \input{introduction.tex}

  \input{overview.tex}
  
  \input{implementation.tex}

  \input{evaluation.tex}

  \input{related_work.tex}

  \input{conclusion.tex}

  \acks This material is based upon work supported by the National
  Science Foundation under Grants No. CCF-1012195 and
  CCF-1439008. Charlie Curtsinger was supported by a Google PhD
  Research Fellowship. The authors thank Dan Barowy, Steve Freund,
  Emma Tosch, John Vilk, and our shepherd Tim Harris for their
  feedback and helpful comments.
  
  \bibliographystyle{abbrv}
  \bibliography{causal,emery}

\end{document}

%% file: abstract.tex
\begin{abstract}

Improving performance is a central concern for software
developers. To locate optimization opportunities, developers rely on
software profilers. However, these profilers only report where
programs spent their time: optimizing that code may have no impact on
performance. Past profilers thus both waste developer time and make it
difficult for them to uncover significant optimization opportunities.

This paper introduces \emph{causal profiling}.  Unlike past profiling
approaches, causal profiling indicates exactly where programmers
should focus their optimization efforts, and quantifies their
potential impact. Causal profiling works by running
\emph{performance experiments} during program execution. Each
experiment calculates the impact of any potential optimization by
\emph{virtually speeding} up code: inserting pauses that slow down all
other code running concurrently. The key insight is that this
slowdown has the same \emph{relative} effect as running that line faster, thus
``virtually'' speeding it up.

We present \coz{}, a causal profiler, which we evaluate on a range of
highly-tuned applications: Memcached, SQLite, and the PARSEC benchmark
suite. \coz{} identifies previously unknown optimization opportunities
that are both significant and targeted. Guided by \coz{}, we improve
the performance of Memcached by 9\%, SQLite by 25\%, and accelerate
six PARSEC applications by as much as 68\%; in most cases, these
optimizations involve modifying under 10 lines of code.

\end{abstract}

%% file: introduction.tex
\section{Introduction}
\label{sec:introduction}


Improving performance is a central concern for software developers.
While compiler optimizations are of some assistance, they often do not
have enough of an impact on performance to meet programmers'
demands~\cite{stabilizer:asplos13}.  
Programmers seeking to increase
the throughput or responsiveness of their applications thus must
resort to manual performance tuning.

Manually inspecting a program to find optimization opportunities is impractical, 
so developers use profilers.
Conventional profilers rank code by its contribution to total execution time.  Prominent examples include oprofile, perf, and gprof~\cite{oprofile, perf_wiki,
  DBLP:conf/sigplan/GrahamKM82}.
Unfortunately, even when a profiler accurately reports where a program
spends its time, this information can lead programmers astray.
Code that runs for a long time is not necessarily a good choice for
optimization.  For example, optimizing code that draws a loading
animation during a file download will not make the program run
faster, even though this code runs just as long as the download.


\begin{figure}[!b]
  \centering\small\textbf{\texttt{example.cpp}}
\begin{lstlisting}
void a() { // ~6.7 seconds
  for(volatile size_t x=0; x<2000000000; x++) {}
}
void b() { // ~6.4 seconds
  for(volatile size_t y=0; y<1900000000; y++) {}
}
int main() {
  // Spawn both threads and wait for them.
  thread a_thread(a), b_thread(b);
  a_thread.join(); b_thread.join();
}
\end{lstlisting}
  \caption{
    A simple multithreaded program that illustrates the shortcomings of existing profilers.
    Optimizing $f_a$ will improve performance by no more than 4.5\%, while optimizing $f_b$ would have no effect on performance.
    \label{fig:simple_example}
  }
\end{figure}


\begin{figure*}[!t]
  \centering
  \begin{subfigure}[t]{0.475\textwidth}
    \centering\small\textbf{Conventional Profile for \texttt{example.cpp}}
\begin{lstlisting}[numbers=none]
    %  cumulative     self            self    total           
 time     seconds  seconds  calls  Ts/call  Ts/call  name    
$\textbf{55.20}$        7.20     7.20      1                     a()
$\textbf{45.19}$       13.09     5.89      1                     b()
 
% time   self   children   called    name
                                     <spontaneous>
  55.0   7.20      0.00              a()
--------------------------------------------------
                                     <spontaneous>
  45.0   5.89      0.00              b()
\end{lstlisting}
    \caption{
      A conventional profile for \texttt{example.cpp}, collected with gprof
    }
    \label{fig:simple_gprof}
  \end{subfigure}
  \hfill
  \begin{subfigure}[t]{0.475\textwidth}
    \centering\small\textbf{~~~~~~~~~~~Causal Profile For \texttt{example.cpp}}\\[0.2em]
    \includegraphics{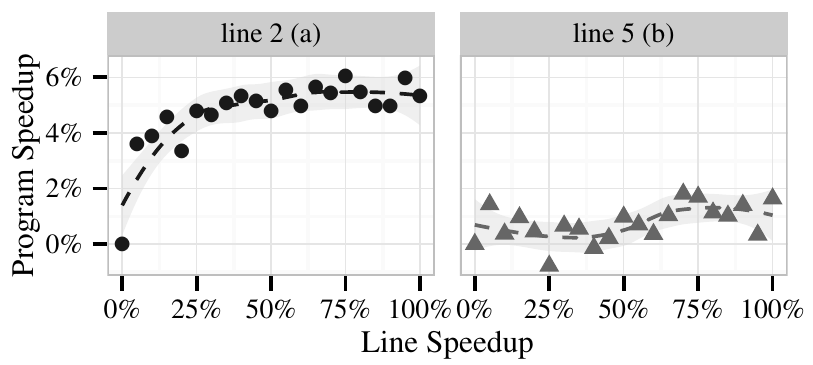}
    \caption{
      Causal profile for \texttt{example.cpp}
      \label{fig:toy_coz}
    }
  \end{subfigure}
  \vspace{0.5em}
  \caption{
    The gprof and causal profiles for the code in Figure~\ref{fig:simple_example}.
    In the causal profile, the y-axis shows the program speedup that would be achieved by speeding up each line of code by the percentage on the x-axis.
    The gray area shows standard error.
    Gprof reports that $f_a$ and $f_b$ comprise similar fractions of total runtime, but optimizing $f_a$ will improve performance by at most 4.5\%, and optimizing $f_b$ would have no effect on performance.
    The causal profile predicts both outcomes within 0.5\%.
  }
\end{figure*}

This phenomenon is not limited to I/O operations.
Figure~\ref{fig:simple_example} shows a simple program that
illustrates the shortcomings of existing profilers, along with its
gprof profile in Figure~\ref{fig:simple_gprof}.  This program spawns
two threads, which invoke functions $f_a$ and $f_b$
respectively.  Most profilers will report that these functions
comprise roughly half of the total execution time.  Other profilers
may report that $f_a$ is on the critical path, or
that the main thread spends roughly equal time waiting for
$f_a$ and $f_b$~\cite{intel_parallel_studio}.  While accurate, all
of this information is potentially misleading.  Optimizing $f_a$ away entirely will only speed up the program by 4.5\% because $f_b$ becomes the new critical path.

Conventional profilers do not report the potential impact of
optimizations; developers are left to make these predictions based on
their understanding of the program.  While these predictions may be
easy for programs as simple as the one in
Figure~\ref{fig:simple_example}, accurately predicting the effect
of a proposed optimization is nearly impossible for programmers
attempting to optimize large applications.


This paper introduces \textbf{\emph{causal profiling}}, an approach
that accurately and precisely indicates where programmers should focus
their optimization efforts, and quantifies their potential impact.
Figure~\ref{fig:toy_coz} shows the results of running \coz{}, our
prototype causal profiler.  This profile plots the hypothetical
speedup of a line of code (x-axis) versus its impact on execution time
(y-axis).  The graph correctly shows that optimizing either $f_a$
or $f_b$ in isolation would have little effect.

A causal profiler conducts a series of \emph{performance experiments} to empirically observe the effect of a potential optimization.
Of course it is not possible to automatically speed up any line of code by an arbitrary amount.
Instead, a causal profiler uses the novel technique of \emph{virtual speedups} to mimic the effect of optimizing a specific line of code by a fixed amount.
A line is virtually sped up by inserting pauses to slow all other threads each time the line runs.
The key insight is that this slowdown has the same relative effect as running that line faster, thus ``virtually'' speeding it up.
Figure~\ref{fig:virtual-speedup} shows the equivalence of virtual and actual speedups.

Each performance experiment measures the effect of virtually speeding up one line by a specific amount.
By conducting many performance experiments over the range of virtual speedup from between 0\% (no change) and 100\% (the line is completely eliminated), a causal profiler can predict the effect of any potential optimization on a program's performance.

Causal profiling further departs from conventional profiling by making it possible to view the effect of optimizations on both \emph{throughput} and \emph{latency}.
To profile throughput, developers specify a \textbf{\emph{progress point}}, indicating a line in the code that corresponds to the end of a unit of work.
For example, a progress point could be the point at which a transaction concludes, when a web page finishes rendering, or when a query completes.
A causal profiler then measures the rate of visits to each progress point to determine any potential optimization's effect on throughput.
To profile latency, programmers instead place two progress points that correspond to the start and end of an event of interest, such as when a transaction begins and completes.
A causal profiler then reports the effect of potential optimizations on the average latency between those two progress points.


To demonstrate the effectiveness of causal profiling, we have developed \coz{}, a causal profiler for Linux.
We show that \coz{} imposes low execution time overhead (mean: 17\%, min: 0.1\%, max: 65\%), making it substantially faster than gprof (up to $6\times$ overhead).

We show that causal profiling accurately predicts optimization opportunities, and that it is effective at guiding optimization efforts.
We apply \coz{} to Memcached, SQLite, and the extensively studied PARSEC benchmark
suite. 
Guided by \coz{}'s output, we optimized the performance of Memcached by 9\%, SQLite by 25\%, and six PARSEC applications by as much as 68\%.
These optimizations typically involved modifying under 10 lines of code.
When possible to accurately measure the size of our optimizations on the line(s)
identified by \coz{}, we compare the observed performance improvements
to \coz{}'s predictions: in each case, we find that the real effect of
our optimization matches \coz{}'s prediction.

\vspace{-0.2em}
\subsection*{Contributions}
This paper makes the following contributions:
\begin{enumerate}

  \item It presents \textbf{causal profiling}, which identifies code where optimizations will have the largest impact. Using \emph{virtual speedups} and \emph{progress points}, causal profiling directly measures the effect of potential optimizations on both throughput and latency ($\S$\ref{sec:overview}).
  
  \item It presents \textbf{\coz{}}, a causal profiler that works on unmodified Linux binaries.
  It describes \coz{}'s implementation ($\S$\ref{sec:implementation}), and demonstrates its efficiency and effectiveness at identifying optimization opportunities ($\S$\ref{sec:evaluation}).

\end{enumerate}


%% file: overview.tex
\section{Causal Profiling Overview}
\label{sec:overview}

This section describes the major steps in collecting, processing, and interpreting a causal profile with \coz{}, our prototype causal profiler.

\paragraph{Profiler startup.}
A user invokes \coz{} using a command of the form \texttt{coz run --- <program> <args>}. At the beginning of the program's execution, \coz{} collects debug information for the executable and all loaded libraries. Users may specify file and binary scope, which restricts \coz{}'s experiments to speedups in the specified files. By default, \coz{} will consider speedups in any source file from the main executable. \coz{} builds a map from instructions to source lines using the program's debug information and the specified scope. Once the source map is constructed, \coz{} creates a profiler thread and resumes normal execution.

\paragraph{Experiment initialization.}
\coz{}'s profiler thread begins an experiment by selecting a line to virtually speed up, and a randomly-chosen percent speedup.
Both parameters must be selected randomly; any systematic method of exploring lines or speedups could lead to systematic bias in profile results.
One might assume that \coz{} could exclude lines or virtual speedup amounts that have not shown a performance effect early in previous experiments, but prioritizing experiments based on past results would prevent \coz{} from identifying an important line if its performance only matters after some warmup period.
Once a line and speedup have been selected, the profiler thread saves the number of visits to each progress point and begins the experiment.

\paragraph{Applying a virtual speedup.}
Every time the profiled program creates a thread, \coz{} begins sampling the instruction pointer from this thread. \coz{} processes samples within each thread to implement a sampling version of virtual speedups. In Section~\ref{sec:implementation_virtual_speedups}, we show the equivalence between the virtual speedup mechanism shown in Figure~\ref{fig:virtual-speedup} and the sampling approach used by \coz{}. Every time a sample is available, a thread checks whether the sample falls in the line of code selected for virtual speedup. If so, it forces other threads to pause. This process continues until the profiler thread indicates that the experiment has completed.

\paragraph{Ending an experiment.}
\coz{} ends the experiment after a pre-determined time has elapsed.
If there were too few visits to progress points during the
experiment---five is the default minimum---\coz{} doubles the
experiment time for the rest of the execution.  Once the experiment
has completed, the profiler thread logs the results of the experiment,
including the effective duration of the experiment (runtime minus the
total inserted delay), the selected line and speedup, and the number
of visits to all progress points.  Before beginning the next
experiment, \coz{} will pause for a brief cooloff period to allow any
remaining samples to be processed before the next experiment begins.


\input{virtual_speedup.tex}


\paragraph{Producing a causal profile.}
After an application has been profiled with \coz{}, the results of all the performance experiments can be combined to produce a causal profile.
Each experiment has two independent variables: the line chosen for virtual speedup and the amount of virtual speedup.
\coz{} records the dependent variable, the rate of visits to each progress point, in two numbers: the total number of visits to each progress point and the effective duration of the experiment (the real runtime minus the total length of all pauses).
Experiments with the same independent variables can be combined by adding the progress point visits and experiment durations.

Once experiments have been combined, \coz{} groups experiments by the line that was virtually sped up.
Any lines that do not have a measurement of 0\% virtual speedup are discarded;
without this baseline measurement we cannot compute a percent speedup relative to the original program.
Measuring this baseline separately for each line guarantees that any line-dependent overhead from virtual speedups, such as the additional cross-thread communication required to insert delays when a frequently-executed line runs, will not skew profile results.
By default, \coz{} will also discard any lines with fewer than 5 different virtual speedup amounts (a plot that only shows the effect of a 75\% virtual speedup is not particularly useful).
Finally, we compute the percent program speedup for each grouped experiment as the percent change in rate of visits to each progress point over the baseline (virtual speedup of 0\%).
\coz{} then plots the resulting table of line and program speedups for each line, producing the profile graphs shown in this paper.

\paragraph{Interpreting a causal profile.}
Once causal profile graphs have been generated, it is up to the user to interpret them and make an educated choice about which lines may be possible to optimize.
To help the user identify important lines, \coz{} sorts the graphs by the slope of their linear regression.
Steep upward slopes indicate a line where optimizations will generally have a positive impact, while a flat line indicates that optimizing this line will not improve program performance.
\coz{} also finds lines with a steep \emph{downward} slope, meaning any optimization to this line will actually hurt performance.
This downward sloping profile is a strong indication of contention;
the line that was virtually sped up interferes with the program's critical path, and optimizing this line increases the amount of interference.
This phenomenon is surprisingly common, and can often result in significant optimization opportunities.
In our evaluation we identify and fix contention issues in three applications: \texttt{fluidanimate}, \texttt{streamcluster}, and \texttt{memcached}, resulting in speedups of 37.5\%, 68.4\%, and 9.4\% respectively.

%% file: virtual_speedup.tex
\begin{figure}[!t]
  \centering\small\textbf{Illustration of Virtual Speedup}
  \includegraphics{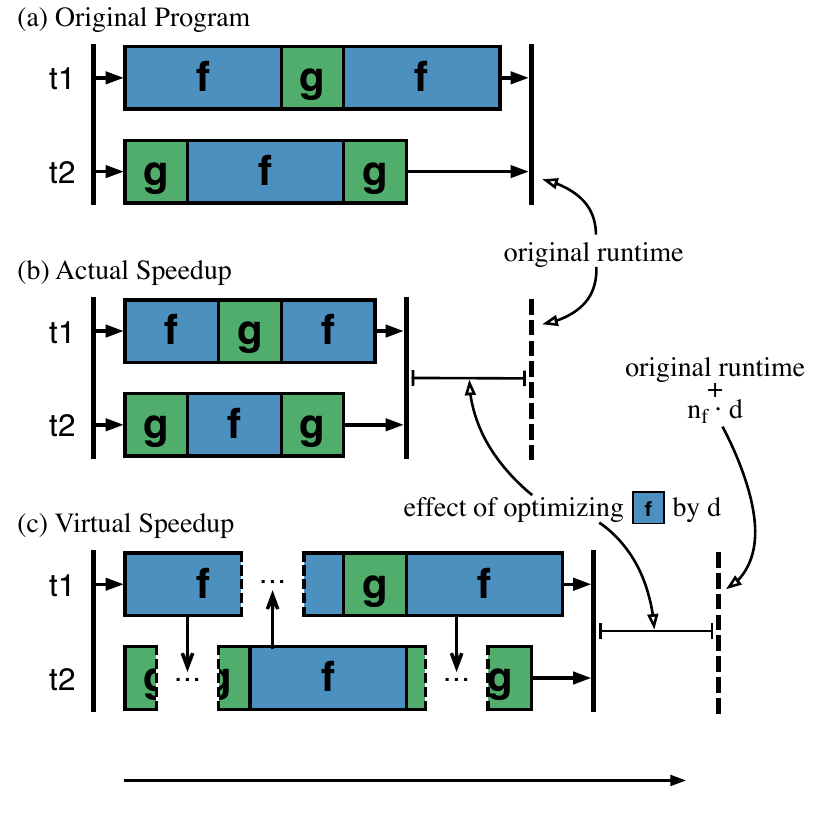}
  \caption{
    An illustration of virtual speedup:
    (a) shows the original execution of two threads running functions \texttt{f} and \texttt{g};
    (b) shows the effect of a \emph{actually} speeding up \texttt{f} by 40\%;
    (c) shows the effect of \emph{virtually} speeding up \texttt{f} by 40\%.
    Each time \texttt{f} runs in one thread, all other threads pause for 40\% of \texttt{f}'s original execution time (shown as ellipsis).
    The difference between the runtime in (c) and the original runtime plus $n_f \cdot d$---the number of times \texttt{f} ran times the delay size---is the same as the effect of actually optimizing \texttt{f}.
    \label{fig:virtual-speedup}
  }
\end{figure}

%% file: implementation.tex
\section{Implementation}
\label{sec:implementation}

This section describes \coz{}'s basic functionality and implementation.
We briefly discuss the core mechanisms required to support profiling unmodified Linux x86-64 executables, along with implementation details for each of the key components of a causal profiler: performance experiments, progress points, and virtual speedups.

\subsection{Core Mechanisms}
\coz{} uses sampling to implement both virtual speedups and progress points.
When a user starts a program with the \texttt{coz} command, \coz{} injects a profiling runtime library into the program's address space using \texttt{LD\_PRELOAD}.
This runtime library creates a dedicated profiler thread to run performance experiments, but also intercepts each thread startup and shutdown to start and stop sampling in the thread using the \texttt{perf\_even} API.
Using the \texttt{perf\_event} API, \coz{} collects both the current program counter and user-space call stack from each thread every 1ms.
To keep overhead low, \coz{} processes samples in batches of ten by default (every 10ms).
Processing samples more frequently is unlikely to improve accuracy, as the additional overhead would distort program execution.

\paragraph{Attributing samples to source locations.}
\coz{} uses DWARF debug information to map sampled program counter values to source locations.
The profiled program does not need to contain DWARF line information;
\coz{} will use the same search procedure as GDB to locate external debug information if necessary~\cite{gdb_manual}.
Note that debug information is available even for optimized code, and most Linux distributions offer packages that include this information for common libraries.

By default, \coz{} will only collect debug information for the main executable.
This means \coz{} will only test potential optimizations in the main program's source files.
Users can specify a source scope to control which source files \coz{} will select lines from to evaluate potential optimizations.
Likewise, users can specify a binary scope to control which executables and libraries will be profiled.
Users should use these scope options to specify exactly which code they are willing or able to change to improve their program's performance.

\subsection{Performance Experiment Implementation}
\label{sec:coz_experiment_init}

\coz{} uses a dedicated profiler thread to coordinate performance experiments.
This thread is responsible for selecting a line to virtually speed up, selecting the size of the virtual speedup, measuring the effect of the virtual speedup on progress points, and writing profiler output.

\paragraph{Starting a performance experiment.}
A single profiler thread, created during program initialization, coordinates performance experiments. 
Before an experiment can begin, the profiler selects a source line to virtually speed up.
To do this, all program threads sample their instruction pointers and map these addresses to source lines.
The first thread to sample a source line that falls within the specified profiling scope sets this as the line selected for virtual speedup.

Once the profiler receives a valid line from one of the program's threads, it chooses a random virtual speedup between 0\% and 100\%, in multiples of 5\%.
For any given virtual speedup, the effect on program performance is $1 - \frac{p_s}{p_0}$, where $p_0$ is the period between progress point visits with no virtual speedup, and $p_s$ is the same period measured with some virtual speedup $s$.
Because $p_0$ is required to compute program speedup for every $p_s$, a virtual speedup of 0 is selected with 50\% probability.
The remaining 50\% is distributed evenly over the other virtual speedup amounts.

Lines for virtual speedup must be selected randomly to prevent bias in the results of performance experiments.
A seemingly reasonably (but invalid) approach would be to begin conducting performance experiments with small virtual speedups, gradually increasing the speedup until it no longer has an effect on program performance.
However, this approach may both over- and under-state the impact of optimizing a particular line if its impact varies over time.

For example, a line that has no performance impact during a program's initialization would not be measured later in execution, when optimizing it could have significant performance benefit.
Conversely, a line that only affects performance during initialization would have exaggerated performance impact unless future experiments re-evaluate virtual speedup values for this line during normal execution.
Any systematic approach to exploring the space of virtual speedup values could potentially lead to systematic bias in the profile output.

Once a line and speedup amount have been selected, \coz{} saves the current values of all progress point counters and begins the performance experiment.

\paragraph{Running a performance experiment.}
Once a performance experiment has started, each of the program's threads processes samples and inserts delays to perform virtual speedups.
After the pre-determined experiment time has elapsed, the profiler thread logs the end of the experiment, including the current time, the number and size of delays inserted for virtual
speedup, the running count of samples in the selected line, and the values for all progress point counters.
After a performance experiment has finished, \coz{} waits until all samples collected during the current experiment have been processed.
By default, \coz{} will process samples in groups of ten, so this pause time is just ten times the sampling rate of 1ms.
Lengthening this cooloff period will reduce \coz{}'s overhead by inserting fewer delays at the cost of increased profiling time to conduct the same number of performance experiments.


\subsection{Progress Point Implementation}
\label{sec:progress_points}

\coz{} supports three mechanisms for monitoring progress points:
\emph{source-level}, \emph{breakpoint}, and
\emph{sampled}.

\paragraph{Source-level progress points.}
Source-level progress points are the only progress points that
require program modification. To indicate a source-level progress
point, a developer simply inserts the \texttt{COZ\_PROGRESS} macro
in the program's source code at the appropriate location.

\paragraph{Breakpoint progress points.}
Breakpoint progress points are specified at the command line.
\coz{} uses the Linux \texttt{perf\_event} API to set a breakpoint at the first
instruction in a line specified in the profiler arguments.

\paragraph{Sampled progress points.}
Sampled progress points are
specified on the command line.  However, unlike source-level and
breakpoint progress points, sampled progress points do not keep
a count of the number of visits to the progress point. Instead,
sampled progress points count the number of samples that fall
within the specified line. As with virtual speedups, the percent
change in visits to a sampled progress point can be computed
even when exact counts are unknown.

\paragraph{Measuring latency.}
Source-level and breakpoint progress points can also be used to
measure the impact of an optimization on latency rather than
throughput. To measure latency, a developer must specify two progress
points: one at the start of some operation, and the other at the
end. The rate of visits to the starting progress point measures the
arrival rate, and the difference between the counts at the start and
end points tells us how many requests are currently in progress. By
denoting $L$ as the number of requests in progress and $\lambda$ as
the arrival rate, we can solve for the average latency $W$ via
Little's Law, which holds for nearly any queuing system: $L = \lambda
W$~\cite{little2011or}. Rewriting Little's Law, we then compute the average
latency as $L / \lambda$.

Little's Law holds under a wide variety of circumstances, and is
independent of the distributions of the arrival rate and service
time. The key requirement is that Little's Law only holds when the
system is \emph{stable}: the arrival rate cannot exceed the service
rate. Note that all usable systems are stable: if a system is
unstable, its latency will grow without bound since the system will
not be able to keep up with arrivals.


\subsection{Virtual Speedup Implementation}
\label{sec:implementation_virtual_speedups}

A critical component of any causal profiler is the ability to virtually speed up any fragment of code.
A naive implementation of virtual speedups is shown in Figure~\ref{fig:virtual-speedup};
each time the function \texttt{f} runs, all other threads are paused briefly.
If \texttt{f} has an average runtime of $\bar{t_f}$ each time it is called and threads are paused for time $d$ each time \texttt{f} runs, then \texttt{f} has an \emph{effective} average runtime of $\bar{t_f}-d$.

If the \emph{real} runtime of \texttt{f} was $\bar{t_f}-d$, but we forced every thread in the program to pause for time $d$ after \texttt{f} ran (including the thread that just executed \texttt{f}) we would measure the same total runtime as with a virtual speedup.
The only difference between virtual speedup and a real speedup with these additional pauses is that we use the time $d$ to allow one thread to finish executing \texttt{f}.
The pauses inserted for virtual speedup increase the total runtime by $n_f \cdot d$, where $n_f$ is the total number of times \texttt{f} by any thread.
Subtracting $n_f \cdot d$ from the total runtime with virtual speedup gives us the execution time we would measure if \texttt{f} had runtime $t_f - d$.

\paragraph{Implementing virtual speedup with sampling.}
The previous discussion of virtual speedups assumes an implementation where every time a specific line of code executes all other threads instantaneously pause for a very brief time (a fraction of the time require to run a single line).
Unfortunately, this approach would incur prohibitively high overhead that would distort program execution, making the profile useless.
Instead, \coz{} periodically samples the program counter and counts samples that fall in the line selected for virtual speedup.
Then, other threads are delayed proportionally to the number of samples.
The number of samples in the selected line, $s$, is approximately
\begin{equation}  \label{eqn:expected_samples}
  s \approx \frac{n \cdot \bar{t}}{P}
\end{equation}
where $P$ is the period of time between samples, $\bar{t}$ is the
average time required to run the selected line once, and $n$ is the
number of times the selected line is executed.

In our original model of virtual speedups, delaying other
threads by time $d$ each time the selected line is executed has
the effect of shortening this line's runtime by $d$. With sampling,
only some executions of the selected line will result in delays.
The effective runtime of the selected line \emph{when sampled} is
$\bar{t} - d$, while executions of the selected line that are not sampled
simply take time $\bar{t}$. The \emph{effective} average time to run the selected
line is
\begin{equation}
  \bar{t}_e = \frac{ (n - s) \cdot \bar{t} + s \cdot (\bar{t} - d) }{n}.
\end{equation}

\noindent Using \eqref{eqn:expected_samples}, this reduces to
\begin{equation}
  \bar{t}_e = \frac{ n \cdot \bar{t} \cdot (1 - \frac{\bar{t}}{P}) + \frac{n \cdot \bar{t}}{P} \cdot (\bar{t} - d) }{ n }
  = \bar{t} \cdot (1 - \frac{d}{P})
\end{equation}

\noindent The relative difference between $t$ and $\bar{t}_e$, the amount of
virtual speedup, is simply
\begin{equation}
  \Delta \bar{t} = 1 - \frac{ \bar{t}_e }{ \bar{t} } = \frac{d}{P}.
\end{equation}

This result lets \coz{} virtually speed up selected lines by a
specific amount without instrumentation. Inserting a delay that is
one quarter of the sampling period will virtually speed up the selected line by
25\%.

\paragraph{Pausing other threads.}
When one thread receives a sample in the line selected for virtual speedup, all other threads must pause.
Rather than using POSIX signals, which would have prohibitively high overhead, \coz{} controls inter-thread pausing using counters.
The first counter, shared by all threads, records the number of times each thread should have paused so far.
Each thread has a local counter of the number of times that thread has already paused.
Whenever a thread's local count of pauses is less than the number of required pauses in the global counter, a thread must pause (and increment its local counter).
To signal all other threads to pause, a thread simply increments both the global counter and its own local counter.
Every thread checks if pauses are required after processing its own samples.

\paragraph{Ensuring accurate timing.} \coz{} uses
the \texttt{nanosleep} POSIX function to insert delays. This function
only guarantees that the thread will pause for \emph{at least} the
requested time, but the pause may be longer than
requested. \coz{} tracks any excess pause time, which is
subtracted from future pauses.

\paragraph{Thread creation.}
To start sampling and adjust delays, \coz{} interposes on the
\texttt{pthread\_create} function. \coz{} first initiates
\texttt{perf\_event} sampling in the new thread. It then inherits the
parent thread's local delay count;
any previously inserted delays to the parent thread also delayed the creation of the new thread.

\subsubsection{Handling Suspended Threads}
\coz{} only collects samples and inserts delays in a thread while that thread is actually executing.
This means that required delays will accumulate in a thread while it is suspended.
When a thread is suspended on a blocking I/O operation, this is the desired behavior;
pausing the thread while it is already suspended on I/O would not delay the thread's progress.
\coz{} simply adds these delays after the thread unblocks.

However, a thread can also be suspended while waiting for a mutex or other POSIX synchronization primitive.
As with blocking I/O, required delays will accumulate while the thread is suspended, but \coz{} may not need to insert all of these delays when the thread resumes.
When one thread resumes after waiting for a mutex, another thread must have unlocked that mutex.
If the unlocking thread has executed all the required delays, then the blocked thread has effectively already been delayed;
it should not insert any additional delays after unblocking.

To correctly handle suspended threads, a causal profiler must follow a simple rule:
If a suspended thread resumes execution because of another thread, the suspended thread should be ``credited'' for any delays inserted in the thread responsible for waking it up.
Otherwise, the thread should insert all the necessary delays that accumulated during the time the thread was suspended.
To simplify the implementation of this policy, \coz{} forces a thread to execute all required delays before it does anything that could block that thread (see Table~\ref{tab:wait_threads}) or wake a suspended thread (shown in Table~\ref{tab:wake_threads}).
This means that any resumed thread can skip any required delays after returning from a call which may have blocked the thread.
Note that this special handling is only required for operations that can suspend a thread.
\coz{} can accommodate programs with ad-hoc synchronization that does not suspend threads with no special handling.

\input{unblocking_table.tex}

\input{blocking_table.tex}

\subsubsection{Attributing Samples to Source Lines}
Samples are attributed to source lines using the source map constructed at
startup. When a sample does not fall in any in-scope source line, the profiler
walks the sampled callchain to find the first in-scope address. This has
the effect of attributing all out-of-scope execution to the last in-scope
callsite responsible. For example, a program may call \texttt{printf}, which
calls \texttt{vfprintf}, which in turn calls \texttt{strlen}. Any samples
collected during this chain of calls will be attributed to the source line that
issues the original \texttt{printf} call.

\subsubsection{Optimization: Minimizing Delays}

If every thread executes the selected line, forcing each thread to
delay $\text{num\_threads}-1$ times unnecessarily slows execution. If
all but one thread executes the selected line, only that thread needs
to pause. The invariant that must be preserved is the following: for
each thread, the number of pauses plus the number of samples in the
selected line must be equal. When a sample falls
in the selected line, \coz{} increments only the local delay count. If
the local delay count is still less than the global delay count after
processing all available samples, \coz{} inserts pauses. If the local
delay count is larger than global delay count, the thread increases
the global delay count.


\subsubsection*{Adjusting for phases}
\coz{} randomly selects a recently-executed line of code at the start
of each performance experiment. This increases the likelihood that
experiments will yield useful information---a virtual speedup would
have no effect on lines that never run---but could bias results for
programs with phases. 

If a program runs in phases, optimizing a line will not have any
effect on progress rate during periods when the line is not being
run. However, \coz{} will not run performance experiments for the line
during these periods because only currently-executing lines are
selected. If left uncorrected, this bias would lead \coz{} to
overstate the effect of optimizing lines that run in phases.

To eliminate this bias, we break the program's execution into two
logical phases: phase A, during which the selected line runs, and
phase B, when it does not. These phases need not be contiguous. The
total runtime $T = t_A + t_B$ is the sum of the durations of the two
phases.  The average progress rate during the entire execution is:

\begin{equation} \label{eqn:P}
  P = \frac{T}{N} = \frac{t_A+t_B}{N}.
\end{equation}

\coz{} collects samples during the entire execution, recording
the number of samples in each line. We define $s$ to be the number of
samples in the selected line, of which $s_{obs}$ occur during a
performance experiment with duration $t_{obs}$. The expected number of
samples during the experiment is:

\begin{equation} \label{eqn:t_A}
  \mathbb{E}[s_{obs}] = s \cdot \frac{t_{obs}}{t_A},
  \,\,\,\,\,\,\,\,
  \text{therefore}
  \,\,\,\,\,\,\,\,
  t_A \approx s \cdot \frac{t_{obs}}{s_{obs}}.
\end{equation}

\coz{} measures the effect of a virtual speedup during phase A,
\[
  \Delta p_A = \frac{p_A - {p_A}^\prime}{p_A}
\]
where ${p_A}^\prime$ and $p_A$ are the average progress periods with and without a virtual speedup; this can be rewritten as:

\begin{equation}
  \Delta p_A = \frac{ \frac{t_A}{n_A} - \frac{{t_A}^\prime}{n_A} } { \frac{t_A}{n_A} }
             = \frac{t_A - {t_A}^\prime}{t_A}  \label{eqn:delta_p_A}
\end{equation}
where $n_A$ is the number of progress point visits during phase A.
Using \eqref{eqn:P}, the new value for $P$ with the virtual speedup is
\[
  P^\prime = \frac{{t_A}^\prime + t_B}{N}
\]
and the percent change in $P$ is
\[
  \Delta P = \frac{P - P^\prime}{P} 
           = \frac{ \frac{t_A + t_B}{N} - \frac{{t_A}^\prime + t_B}{N} }{ \frac{T}{N} } 
           = \frac{t_A - {t_A}^\prime}{T}.
\]
Finally, using \eqref{eqn:t_A} and \eqref{eqn:delta_p_A},
\begin{equation} \label{eqn:phase_correction}
  \Delta P = \Delta p_A \frac{t_A}{T}
            \approx \Delta p_A \cdot \frac{t_{obs}}{s_{obs}} \cdot \frac{s}{T}.
\end{equation}

\coz{} multiplies all measured speedups, $\Delta p_A$, by the correction factor $\frac{t_{obs}}{s_{obs}} \cdot \frac{s}{T}$ in its final report.

%% file: unblocking_table.tex
\begin{table}[t]
\small
\centering
\begin{tabular}{ll}
\multicolumn{2}{c}{\textbf{Potentially \emph{unblocking} calls}} \\
\hline
\texttt{pthread\_mutex\_unlock}   & unlock a mutex \\
\texttt{pthread\_cond\_signal}    & wake one waiter on a c.v.\\
\texttt{pthread\_cond\_broadcast} & wake all waiters on c.v. \\
\texttt{pthread\_barrier\_wait}   & wait at a barrier \\
\texttt{pthread\_kill}            & send signal to a thread \\
\texttt{pthread\_exit}            & terminate this thread \\
\end{tabular}
\caption{\coz{} intercepts POSIX functions that could wake a blocked thread. To ensure correctness of virtual speedups, \coz{} forces threads to execute any unconsumed delays before invoking any of these functions and potentially waking another thread.\label{tab:wake_threads}}
  \vspace{0.7em}
\end{table}

%% file: blocking_table.tex
\begin{table}[t]
\small
\centering
\begin{tabular}{ll}
\multicolumn{2}{c}{\textbf{Potentially \emph{blocking} calls}} \\
\hline
\texttt{pthread\_mutex\_lock}   & lock a mutex \\
\texttt{pthread\_cond\_wait}    & wait on a condition variable \\
\texttt{pthread\_barrier\_wait} & wait at a barrier \\
\texttt{pthread\_join}          & wait for a thread to complete \\
\texttt{sigwait}                & wait for a signal \\
\texttt{sigwaitinfo}            & wait for a signal \\
\texttt{sigtimedwait}           & wait for a signal (with timeout) \\
\texttt{sigsuspend}             & wait for a signal \\
\end{tabular}
\caption{\coz{} intercepts POSIX functions that could block waiting for a thread, instrumenting them to update delay counts before and after blocking.\label{tab:wait_threads}}
\end{table}

%% file: evaluation.tex
\input{optimizations_table.tex}

\section{Evaluation}
\label{sec:evaluation}
Our evaluation answers the following questions:
(1) Does causal profiling enable effective performance tuning?
(2) Are \coz{}'s performance predictions accurate?
(3) Is \coz{}'s overhead low enough to be practical?

\subsection{Experimental Setup}

We perform all experiments on a 64 core, four socket AMD Opteron
machine with 60GB of memory, running Linux 3.14 with no modifications.
All applications are compiled using GCC version 4.9.1 at the
\texttt{-O3} optimization level and debug information generated with
\texttt{-g}.  We disable frame pointer elimination with the
\texttt{-fno-omit-frame-pointer} flag so the Linux can collect
accurate call stacks with each sample.  \coz{} is run with the default
sampling period of 1ms, with sample processing set to occur after every 10 samples.  Each
performance experiment runs with a cooling-off period of 10ms after each experiment to allow any remaining samples to be processed before the next experiment begins.  Due to space limitations, we
only profile throughput (and not latency) in this evaluation.

\subsection{Effectiveness}
\label{sec:effectiveness}

We demonstrate causal profiling's effectiveness through case studies.
Using \coz{}, we collect causal profiles for Memcached, SQLite, and the PARSEC benchmark suite.
Using these causal profiles, we were able to make small changes to two of the real applications and six PARSEC benchmarks, resulting in performance improvements as large as 68\%.
Table~\ref{fig:speedup_summary} summarizes the results of our optimization efforts.
We describe our experience using \coz{} below, with three general outcomes:
(1) cases where \coz{} found optimization opportunities that gprof and perf did not (dedup, ferret, and SQLite);
(2) cases where \coz{} identified contention (fluidanimate, streamcluster, and Memcached); and
(3) cases where both \coz{} and a conventional profiler identified the optimization we implemented (blackscholes and swaptions).

\input{dedup.tex}

\input{ferret.tex}

\begin{figure}[!t]
  \centering\small\textbf{~~~~~~~~~Causal Profile for ferret}\\[0.3em]
  \includegraphics{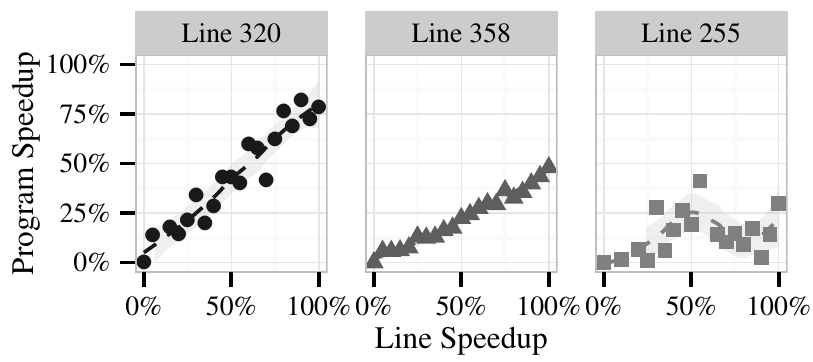}
  \caption{
    \small
    \coz{} output for the unmodified ferret application. The x-axis shows the amount of virtual speedup applied to each line, versus the resulting change in throughput on the y-axis. The top two lines are executed by the indexing and ranking stages; the third line is executed during image segmentation.
    \label{fig:ferret}
  }
\end{figure}

\input{sqlite.tex}


\input{fluidanimate.tex}

\input{streamcluster.tex}

\begin{figure}[!t]
  \centering\small\textbf{~~~~~~~~~Causal Profile for fluidanimate}\\[0.3em]
  \includegraphics{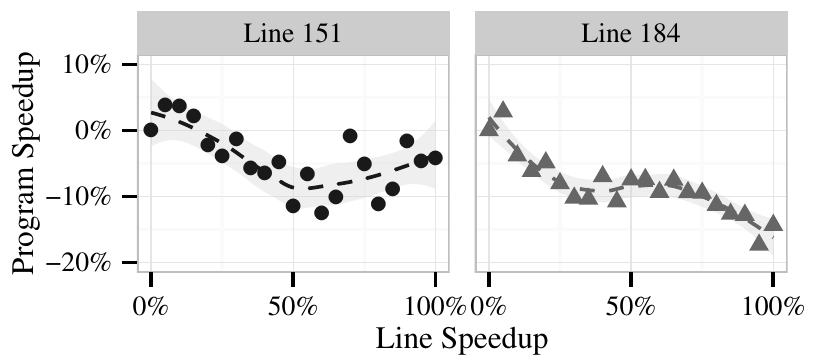}
  \caption{
    \small
    \coz{} output for fluidanimate, prior to optimization.
    \coz{} finds evidence of contention in two lines in \texttt{parsec\_barrier.cpp}, the custom barrier implementation used by both fluidanimate and streamcluster.
    This causal profile reports that optimizing either line will slow down the application, not speed it up.
    These lines precede calls to \texttt{pthread\_mutex\_trylock} on a contended mutex.
    Optimizing this code would increase contention on the mutex and interfere with the application's progress.
    Replacing this inefficient barrier implementation sped up fluidanimate and streamcluster by $37.5\%$ and $68.4\%$ respectively.
    \label{fig:fluidanimate}
  }
\end{figure}

\input{memcached.tex}


\input{blackscholes.tex}
\input{swaptions.tex}



\paragraph{Effectiveness Summary.}
Our case studies confirm that \coz{} is effective at identifying optimization opportunities and guiding performance tuning.
In every case, the information \coz{} provided led us directly to the optimization we implemented.
In most cases, \coz{} identified around 20 lines of interest, with as many as 50 for larger programs (Memcached and x264).
\coz{} identified optimization opportunities in all of the PARSEC benchmarks, but some required more invasive changes that are out of scope for this paper.
Table~\ref{table:benchmarks} summarizes our findings for the remaining PARSEC benchmarks.
We have submitted patches to the developers of all the applications we optimized.

\input{benchmarks_table.tex}

\subsection{Accuracy}
For most of the optimizations described above, it is not possible to
quantify the effect our optimization had on the specific lines that
\coz{} identified. However, for two of our case studies---ferret and
dedup---we can directly compute the effect our optimization had on the
line \coz{} identified and compare the resulting speedup to \coz{}'s
predictions. Our results show that \coz{}'s predictions
are highly accurate.

To optimize ferret, we increased the number of threads for the
indexing stage from $16$ to $22$, which increases the throughput of
line 320 by $27\%$.  \coz{} predicted that this improvement would
result in a $21.4\%$ program speedup, which is nearly the same as the
$21.2\%$ we observe.

For dedup, \coz{} identified the top of the \texttt{while} loop that traverses a hash bucket's linked list.
By replacing the degenerate hash function, we reduced the average number of elements in each hash bucket from $76.7$ to just $2.09$.
This change reduces the number of iterations from $77.7$ to $3.09$ (accounting for the final trip through the loop).
This reduction corresponds to a speedup of the line \coz{} identified by $96\%$.
For this speedup, \coz{} predicted a performance improvement of $9\%$, very close to our observed speedup of $8.95\%$.

\begin{figure}[!t]
  \centering\small\textbf{~~~~~~~~~Overhead of \coz{}}\\[0.3em]
  \includegraphics{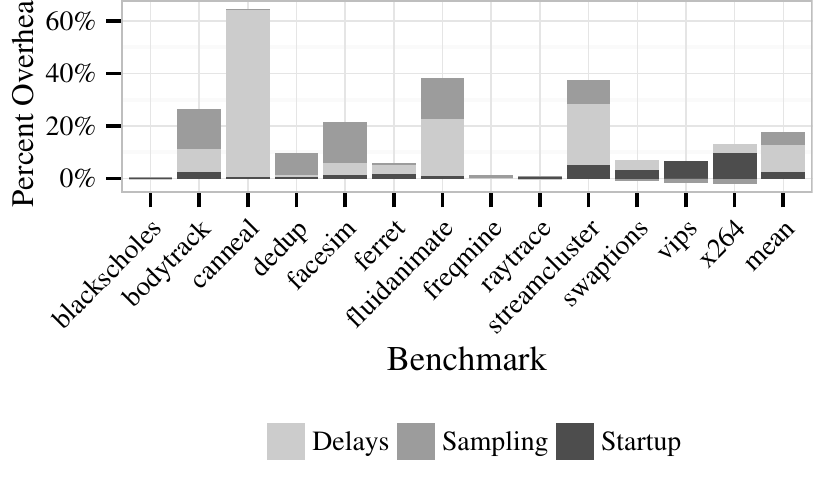}
  \caption{
    Percent overhead for each of \coz{}'s possible sources of overhead. \emph{Delays} are the overhead due to adding delays for virtual speedups, \emph{Sampling} is the cost of collecting and processing samples, and \emph{Startup} is the initial cost of processing debugging information. Note that sampling results in slight performance \emph{improvements} for swaptions, vips, and x264.
  \label{fig:overhead}
  }
\end{figure}

\subsection{Efficiency}

We measure \coz{}'s profiling overhead on the PARSEC benchmarks
running with the native inputs. The sole exception is streamcluster,
where we use the test inputs because execution time was excessive
with the native inputs.

Figure~\ref{fig:overhead} breaks down the total overhead of running \coz{} on each of the PARSEC benchmarks by category.
The average overhead with \coz{} is 17.6\%.
\coz{} collects debug information at startup, which contributes 2.6\% to the average overhead.
Sampling during program execution and attributing these samples to lines using debug information is responsible for 4.8\% of the average overhead.
The remaining overhead (10.2\%) comes from the delays \coz{} inserts to perform virtual speedups.

These results were collected by running each benchmark in four configurations.
First, each program was run without \coz{} to measure a baseline execution time.
In the second configuration, each program was run with \coz{}, but execution terminated immediately after startup work was completed.
Third, programs were run with \coz{} configured to sample the program's execution but not to insert delays (effectively testing only virtual speedups of size zero).
Finally, each program was run with \coz{} fully enabled.
The difference in execution time between each successive configuration give us the startup, sampling, and delay overheads, respectively.

\paragraph{Reducing overhead.}
Most programs have sufficiently long running times (mean: 103s) to amortize the cost of processing debug information, but especially large executables can be expensive to process at startup (\texttt{x264} and \texttt{vips}, for example).
\coz{} could be modified to collect and process debug information lazily to reduce startup overhead.
Sampling overhead comes mainly from starting and stopping sampling with the \texttt{perf\_event} API at thread creation and exit.
This cost could be amortized by sampling globally instead of per-thread, which would require root permissions on most machines.
If the \texttt{perf\_event} API supported sampling all threads in a process this overhead could be eliminated.
Delay overhead, the largest component of \coz{}'s total overhead, could be reduced by allowing programs to execute normally for some time between each experiment.
Increasing the time between experiments would significantly reduce overhead, but a longer profiling run would be required to collect a usable profile.

\paragraph{Efficiency summary.}
\coz{}'s profiling overhead is on average 17.6\% (minimum: 0.1\%,
maximum: 65\%). For all but three of the benchmarks, its overhead was
under 30\%. Given that the widely used gprof profiler can impose
much higher overhead (e.g., $6\times$ for ferret, versus 6\% with
\coz{}), these results confirm that \coz{} has sufficiently low overhead
to be used in practice.

%% file: optimizations_table.tex
\begin{table}[t]
  \centering\small\textbf{Summary of Optimization Results}
  \begin{tabular}{r | l l l}
    \textit{Application}    & \textit{Speedup}      & \textit{Diff Size}  & \textit{LOC}  \\
    \hline
    blackscholes   & $2.56\% \pm 0.41\%$   & $-61$, $+4$         & 342   \\
    dedup          & $8.95\% \pm 0.27\%$   & $-3$, $+3$          & 2,570  \\
    ferret         & $21.27\% \pm 0.17\%$  & $-4$, $+4$          & 5,937  \\
    fluidanimate   & $37.5\% \pm 0.56\%$   & $-1$, $+0$          & 1,015  \\
    streamcluster  & $68.4\% \pm 1.12\%$   & $-1$, $+0$          & 1,779  \\
    swaptions      & $15.8\% \pm 1.10\%$   & $-10$, $+16$        & 970   \\
    \hline
    Memcached      & $9.39\% \pm 0.95\%$   & $-6$, $+2$          & 10,475 \\
    SQLite        & $25.60\% \pm 1.00\%$  & $-7$, $+7$          & 92,635 \\
  \end{tabular}
  \caption{
    All benchmarks were run ten times before and after optimization.
    Standard error for speedup was computed using Efron's bootstrap method, where speedup is defined as $\frac{t_0-t_{opt}}{t_0}$.
    All speedups are statistically significant at the 99.9\% confidence level ($\alpha = 0.001$) using the one-tailed Mann-Whitney U test, which does not rely on any assumptions about the distribution of execution times.
    Lines of code do not include blank or comment-only lines.
    \label{fig:speedup_summary}
  }
\end{table}

%% file: dedup.tex
\begin{figure}[t]
  \centering\small\textbf{~~~~~~~~~Hash Bucket Collisions in dedup}\\[0.3em]
  \includegraphics{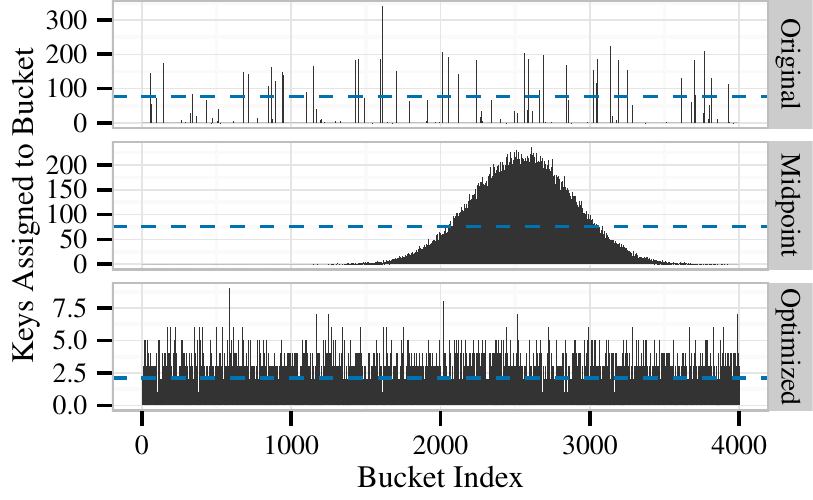}
  \caption{
    In the dedup benchmark, \coz{} identified hash bucket traversal as a bottleneck. 
    These plots show collisions per-bucket before, mid-way through, and after optimization of the dedup benchmark (note different y-axes). The dashed horizontal line shows average collisions per-utilized bucket for each version. Fixing dedup's hash function improved performance by $8.9\%$.
    \label{fig:hashes}
  }
\end{figure}

\subsubsection{Case Study: dedup}

The dedup application performs parallel file compression via deduplication.
This process is divided into three main stages: fine-grained fragmentation, hash computation, and compression.
We placed a progress point immediately after dedup completes compression of a single block of data (\texttt{encoder.c:189}).

\coz{} identifies the source line \texttt{hashtable.c:217} as the best opportunity for optimization.
This code is the top of the \texttt{while} loop in \texttt{hashtable\_search} that traverses the linked list of entries that have been assigned to the same hash bucket.
This suggests that dedup's shared hash table has a significant number of collisions.
Increasing the hash table size had no effect on performance.
This led us to examine dedup's hash function, which could also be responsible for the large number of hash table collisions.
We discovered that dedup's hash function maps keys to just 2.3\% of the available buckets;
over 97\% of buckets were never used during the entire execution.

The original hash function adds characters of the hash table key,
which leads to virtually no high order bits being set.  The resulting
hash output is then passed to a bit shifting procedure intended to
compensate for poor hash functions.  We removed the bit shifting step,
which increased hash table utilization to $54.4\%$.  We then changed
the hash function to bitwise XOR 32 bit chunks of the key.  This
increased hash table utilization to $82.0\%$ and resulted in an
$8.95\% \pm 0.27\%$ performance improvement.  Figure~\ref{fig:hashes}
shows the rate of bucket collisions of the original hash function, the
same hash function without the bit shifting ``improvement'', and our
final hash function.  The entire optimization required changing just
three lines of code.  As with ferret, this result was achieved by one
graduate student who was initially unfamiliar with the code; the
entire profiling and tuning effort took just two hours.

\paragraph{Comparison with gprof.}

We ran both the original and optimized versions of dedup with gprof.
As with ferret, the optimization opportunities identified by \coz{}
were not obvious in gprof's output.  Overall,
\texttt{hashtable\_search} had the largest share of highest execution
time at 14.38\%, but calls to \texttt{hashtable\_search} from the hash
computation stage accounted for just 0.48\% of execution time; Gprof's
call graph actually obscured the importance of this code.  After
optimization, \texttt{hashtable\_search}'s share of execution time
reduced to 1.1\%.

%% file: ferret.tex
\begin{figure}[!t]
\vspace{0.2em}
 \centering \includegraphics[width=3.25in]{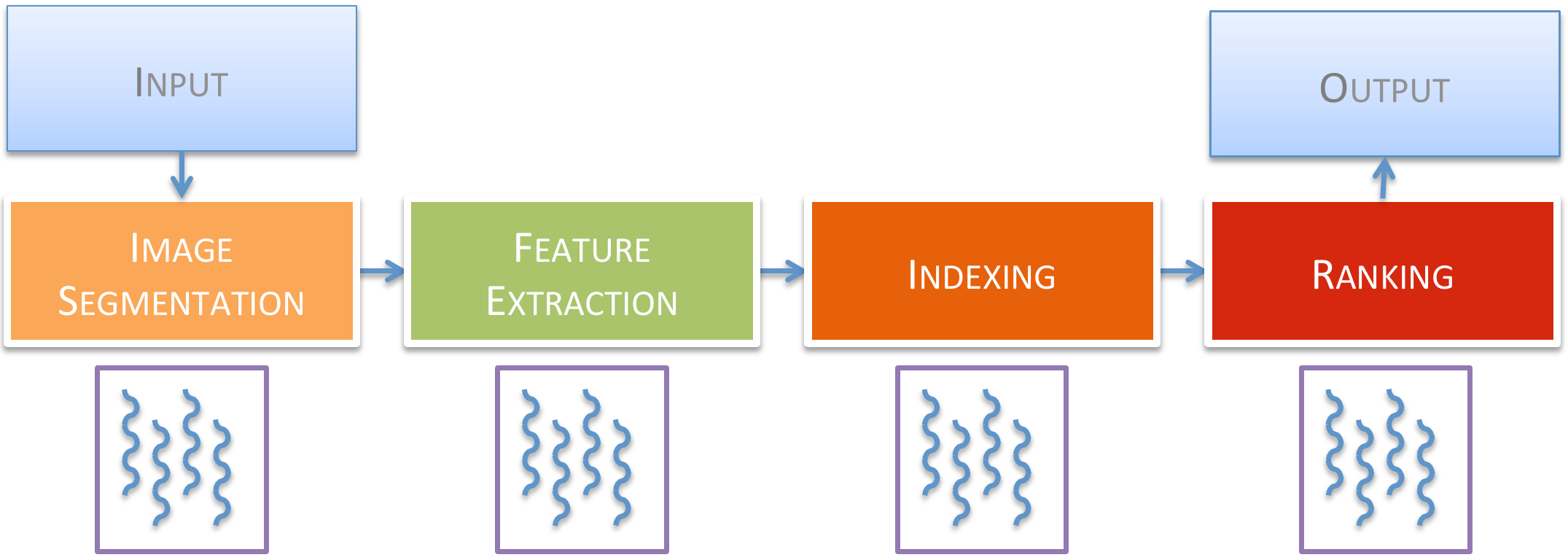}
 \caption{\small Ferret's pipeline. The middle four stages each have an associated thread pool; the input and output stages each consist of one thread. The colors represent the impact on throughput of each stage, as identified by \coz{}: green is low impact, orange is medium impact, and red is high impact.\label{fig:ferret-pipeline}}
\end{figure}

\subsubsection{Case Study: ferret}

The ferret benchmark performs a content-based image similarity search.
Ferret consists of a pipeline with six stages: the first and the last
stages are for input and output.  The middle four stages perform image
segmentation, feature extraction, indexing, and ranking.  Ferret takes
two arguments: an input file and a desired number of threads, which
are divided equally across the four middle stages.  We first inserted
a progress point in the final stage of the image search pipeline to
measure throughput (\texttt{ferret-parallel.c:398}).
We then ran \coz{} with the source scope set to evaluate optimizations only in \texttt{ferret-parallel.c}, rather than across the entire ferret toolkit.

Figure~\ref{fig:ferret} shows the top three lines identified by
\coz{}, using its default ranking metric.  Lines 320 and 358 are calls
to \texttt{cass\_table\_query} from the indexing and ranking stages.
Line 255 is a call to \texttt{image\_segment} in the segmentation
stage.  Figure~\ref{fig:ferret-pipeline} depicts ferret's pipeline
with the associated thread pools (colors indicate \coz{}'s computed
impact on throughput of optimizing these stages).

Because each important line falls in a different pipeline stage, and
because \coz{} did not find any important lines in the queues shared
by adjacent stages, we can easily ``optimize'' a specific line by
shifting threads to that stage.  We modified ferret to let us specify
the number of threads assigned to each stage separately, a four-line
change.

\coz{} did not find any important lines in the feature extraction
stage, so we shifted threads from this stage to the three other main
stages.  After three rounds of profiling and adjusting thread
assignments, we arrived at a final thread allocation of 20, 1, 22, and
21 to segmentation, feature extraction, indexing, and ranking
respectively.  The reallocation of threads led to a $21.27\% \pm
0.17\%$ speedup over the original configuration, using the same number
of threads.


\paragraph{Comparison with gprof.}

We also ran ferret with gprof in both the initial and final
configurations.  Optimization opportunities are not immediately
obvious from that profile.  For example, in the flat profile, the
function \texttt{cass\_table\_query} appears near the bottom of the
ranking, and is tied with 56 other functions for most cumulative time.

Gprof also offers little guidance for optimizing ferret. In fact, its
output was virtually unchanged before and after our optimization,
despite a large performance change.

%% file: sqlite.tex
\begin{figure}[!t]
  \centering\small\textbf{~~~~~~~~~Causal and Perf Profiles for SQLite}\\[0.3em]
  \begin{subfigure}[t]{0.475\textwidth}
    \includegraphics{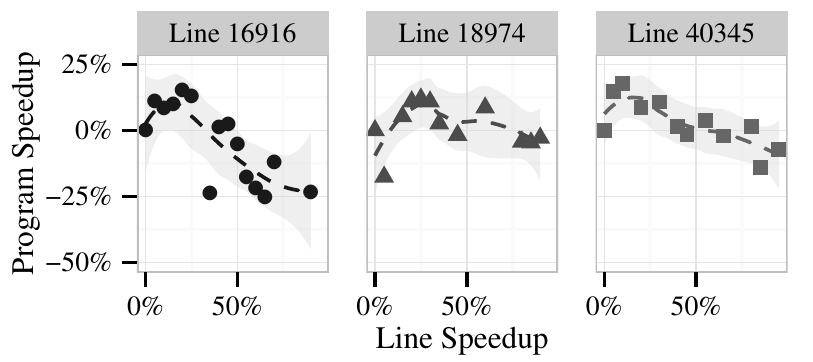}
    \caption{
      \coz{}'s output for SQLite before optimizations.
      \label{fig:sqlite_coz}
      \vspace{0.75em}
    }
  \end{subfigure}
  \begin{subfigure}[t]{0.475\textwidth}
\begin{lstlisting}[numbers=none]
% Runtime    Symbol
  85.55%     _raw_spin_lock
  1.76%      x86_pmu_enable_all
     $\emph{... 30 lines ...}$
  0.10%      rcu_irq_enter
  $\textbf{0.09\%~~~~~~sqlite3MemSize}$
  0.09%      source_load
    $\emph{... 26 lines ...}$
  0.03%      __queue_work
  $\textbf{0.03\%~~~~~~pcache1Fetch}$
  0.03%      kmem_cache_free
  0.03%      update_cfs_rq_blocked_load
  $\textbf{0.03\%~~~~~~pthreadMutexLeave}$
  0.03%      sqlite3MemMalloc
\end{lstlisting}
    \vspace{-0.5em}
    \caption{
      Perf's output for SQLite before optimizations.
      \label{fig:sqlite_perf}
    } 
  \end{subfigure}
  \vspace{0.5em}
  \caption{
    \small
    \coz{} and perf output for SQLite before optimizations.
    The three lines in the causal profile correspond to the function prologues for \texttt{sqlite3MemSize}, \texttt{pthreadMutexLeave}, and \texttt{pcache1Fetch}.
    A small optimization to each of these lines will improve program performance, but beyond about a 25\% speedup, \coz{} predicts that the optimization would actually lead to a slowdown.
    Changing indirect calls into direct calls for these functions improved overall performance by $25.6\% \pm 1.0\%$.
    \label{fig:sqlite}
  }
\end{figure}

\subsubsection{Case Study: SQLite}

The SQLite database library is widely used by many applications to
store relational data.  The embedded database, which can be included
as a single large C file, is used for many applications including
Firefox, Chrome, Safari, Opera, Skype, iTunes, and is a standard
component of Android, iOS, Blackberry 10 OS, and Windows Phone 8.  We
evaluated SQLite performance using a write-intensive parallel
workload, where each thread rapidly inserts rows to its own private
table.  While this benchmark is synthetic, it exposes any
scalability bottlenecks in the database engine itself because all
threads should theoretically operate independently.  We placed a
progress point in the benchmark itself (which is linked with the
database), which executes after each insertion.

\coz{} identified three important optimization opportunities, shown in Figure~\ref{fig:sqlite_coz}.  At
startup, SQLite populates a large number of structs with function
pointers to implementation-specific functions, but most of these
functions are only ever given a default value determined by compile-time options.  The three functions
\coz{} identified unlock a standard pthread mutex, retrieve the next
item from a shared page cache, and get the size of an allocated
object.  These simple functions do very little work, so the overhead
of the indirect function call is relatively high.  Replacing these
indirect calls with direct calls resulted in a $25.60\% \pm 1.00\%$
speedup.

\paragraph{Comparison with conventional profilers.}
Unfortunately, running SQLite with
gprof segfaults immediately.  The application does run with the Linux
perf tool, which reports that the three functions \coz{}
identified account for a total of just $0.15\%$ of total runtime (shown in Figure~\ref{fig:sqlite_perf}). Using perf, a developer
would be misled into thinking that optimizing these functions would be a waste of time.
\coz{} accurately shows that the opposite is true: optimizing these functions
has a dramatic impact on performance.

%% file: fluidanimate.tex
\subsubsection{Case Study: fluidanimate}

The fluidanimate benchmark, also provided by Intel, is a physical
simulation of an incompressible fluid for animation.  The application
spawns worker threads that execute in eight concurrent phases,
separated by a barrier.  We placed a progress point immediately after
the barrier, so it executes each time all threads complete a phase
of the computation.

\coz{} identifies a single modest potential speedup in the thread
creation code, but there was no obvious way to speed up this code.
However, \coz{} also identified two significant points of contention,
indicated by a downward sloping causal profile. 
Figure~\ref{fig:fluidanimate} shows \coz{}'s output for these two lines.
 This result tells us
that optimizing the indicated line of code would actually \emph{slow
  down} the program, rather than speed it up.  Both lines \coz{}
identifies are in a custom barrier implementation, immediately before
entering a loop that repeatedly calls
\texttt{pthread\_mutex\_trylock}.  Removing this spinning from the
barrier would reduce the contention, but it was simpler to replace the
custom barrier with the default \texttt{pthread\_barrier}
implementation.  This one line change led to a $37.5\% \pm 0.56\%$
speedup.

%% file: streamcluster.tex
\subsubsection{Case Study: streamcluster}

The streamcluster benchmark performs online clustering of streaming
data.  As with fluidanimate, worker threads execute in concurrent
phases separated by a custom barrier, where we placed a progress
point.  \coz{} identified a call to a random number generator as a
potential line for optimization.  Replacing this call with a
lightweight random number generator had a modest effect on performance
(\~{}2\% speedup).  As with fluidanimate, \coz{} highlighted the custom
barrier implementation as a major source of contention.  Replacing
this barrier with the default \texttt{pthread\_barrier} led to a
$68.4\% \pm 1.12\%$ speedup.

%% file: memcached.tex
\subsubsection{Case Study: Memcached}

Memcached is a widely-used in-memory caching system.  To evaluate
cache performance, we ran a benchmark ported from the Redis
performance benchmark.  This program spawns 50 parallel clients that
collectively issue 100,000 \texttt{SET} and \texttt{GET} requests for
randomly chosen keys.  We placed a progress point at the end of the
\texttt{process\_command} function, which handles each client request.

Most of the lines \coz{} identifies are cases of contention, with a
characteristic downward-sloping causal profile plot.  One such line
is at the start of \texttt{item\_remove}, which locks an item in the
cache and then decrements its reference count, freeing it if the count
goes to zero.  To reduce lock initialization overhead, Memcached uses
a static array of locks to protect items, where each item selects its
lock using a hash of its key. Consequently, locking any one item can
potentially contend with independent accesses to other items whose
keys happen to hash to the same lock index. Because reference counts
are updated atomically, we can safely remove the lock from this
function, which resulted in a $9.39\% \pm 0.95\%$ speedup.

%% file: blackscholes.tex
\subsubsection{Case Study: blackscholes}

The blackscholes benchmark, provided by Intel, solves the
Black--Scholes differential equation to price a portfolio of stock
options.  We placed a progress point after each thread completes one
round of the iterative approximation to the differential equation
(\texttt{blackscholes.c:259}).  \coz{} identifies many lines in the
\texttt{CNDF} and \texttt{BlkSchlsEqEuroNoDiv} functions that would
have a small impact if optimized. This same code was identified as a
bottleneck by ParaShares~\cite{DBLP:conf/europar/KambadurTK14}; this
is the only optimization we describe here that was previously reported.
This block of code performs the main numerical work of the program,
and uses many temporary variables to break apart the complex
computation.  Manually eliminating common subexpressions and combining
61 piecewise calculations into 4 larger expressions resulted in a
$2.56\% \pm 0.41\%$ program speedup.

%% file: swaptions.tex
\subsubsection{Case Study: swaptions}

The swaptions benchmark is a Monte Carlo pricing algorithm for
swaptions, a type of financial derivative.  Like blackscholes and
fluidanimate, this program was developed by Intel.  We placed a
progress point after each iteration of the main loop executed by worker threads
(\texttt{HJM\_Securities.cpp:99}).

\coz{} identified three significant optimization opportunities, all
inside nested loops over a large multidimensional array.  One of these
loops zeroed out consecutive values.  A second loop filled
part of the same large array with values from a distribution function,
with no obvious opportunities for optimization.  The third nested loop
iterated over the same array again, but traversed the dimensions in an
irregular order.
Reordering these loops and replacing the first loop with a call to \texttt{memset} sped execution by $15.8\% \pm 1.10\%$.

%% file: benchmarks_table.tex
\begin{table}[t]
  \centering\small\textbf{Results for Unoptimized Applications}\\[0.3em]
  \begin{tabular}{r | l l}
    \textit{Benchmark}
      & \textit{Progress Point}
      & \textit{Top Optimization} \\
    \hline
    
    bodytrack
      & \fontsize{6.5pt}{6pt}\texttt{TicketDispenser.h:106}
      & \fontsize{6.5pt}{6pt}\texttt{ParticleFilter.h:262} \\
      
    canneal
      & \fontsize{6.5pt}{6pt}\texttt{annealer\_thread.cpp:87}
      & \fontsize{6.5pt}{6pt}\texttt{netlist\_elem.cpp:82} \\
          
    facesim
      & \fontsize{6.5pt}{6pt}\texttt{taskQDistCommon.c:109}
      & \fontsize{6.5pt}{6pt}\texttt{MATRIX\_3X3.h:136} \\
              
    freqmine
      & \fontsize{6.5pt}{6pt}\texttt{fp\_tree.cpp:383}
      & \fontsize{6.5pt}{6pt}\texttt{fp\_tree.cpp:301} \\

    raytrace
      & \fontsize{6.5pt}{6pt}\texttt{BinnedAllDims\dots:98}
      & \fontsize{6.5pt}{6pt}\texttt{RTEmulatedSSE.hxx:784} \\
      
    vips
      & \fontsize{6.5pt}{6pt}\texttt{threadgroup.c:360}
      & \fontsize{6.5pt}{6pt}\texttt{im\_Lab2LabQ.c:98} \\
      
    x264
      & \fontsize{6.5pt}{6pt}\texttt{encoder.c:1165}
      & \fontsize{6.5pt}{6pt}\texttt{common.c:687} \\

\punt{      
    \hline
    
    redis
      & \fontsize{6.5pt}{6pt}\texttt{networking.c:880}
      & \fontsize{6.5pt}{6pt}\texttt{redis.c:447} \\
}
  \end{tabular}
  \caption{
    The locations of inserted progress points for the remaining PARSEC benchmarks, and the top optimization opportunities that \coz{} identifies.
    The progress point for raytrace was placed on line 98 of \texttt{BinnedAllDimsSaveSpace.cxx}.
    \label{table:benchmarks}
  }
\end{table}

%% file: related_work.tex
\section{Related Work}
\label{sec:related_work}


Causal profiling identifies and quantifies optimization opportunities,
while most past work on profilers has focused on collecting detailed
(though not necessarily actionable) information with low overhead.


\subsection{General-Purpose Profilers}

General-purpose profilers are typically implemented using instrumentation, sampling, or both.
Systems based on sampling (including causal profiling) can arbitrarily reduce \emph{probe effect}, although sampling must be unbiased~\cite{DBLP:conf/pldi/MytkowiczDHS10}.

%
%
The UNIX prof tool and oprofile both use sampling exclusively~\cite{thompson1975unix,oprofile}.
Oprofile can sample using a variety of hardware performance counters, which can be used to identify cache-hostile code, poorly predicted branches, and other hardware bottlenecks. 
Gprof combines instrumentation and sampling to measure execution time~\cite{DBLP:conf/sigplan/GrahamKM82}.
Gprof produces a call graph profile, which counts invocations of functions segregated
by caller.
Cho, Moseley, et al. reduce the overhead of Gprof's call-graph profiling by interleaving instrumented and un-instrumented execution~\cite{DBLP:conf/cgo/ChoMHBM13}.
Path profilers add further detail, counting executions of each path through a procedure, or across procedures~\cite{DBLP:conf/micro/BallL96,DBLP:conf/pldi/AmmonsBL97}.


\subsection{Parallel Profilers}
Past work on parallel profiling has focused on identifying the critical path or bottlenecks, although optimizing the critical path or removing the bottleneck may not significantly improve program performance.

\paragraph{Critical path profiling.}
IPS uses traces from message-passing programs to identify the critical path, and reports the amount of time each procedure contributes to the critical path~\cite{DBLP:conf/icdcs/MillerY87}.
IPS-2 extends this approach with limited support for shared memory parallelism~\cite{DBLP:journals/tse/YangM89, DBLP:journals/tpds/MillerCHKLT90}.
Other critical path profilers rely on languages with first-class threads and synchronization to identify the critical path~\cite{DBLP:conf/pdp/HillJSV98, DBLP:conf/ipps/OyamaTY00, DBLP:conf/sc/SzebenyiWW09}.
Identifying the critical path helps developers find code where optimizations will have some impact, but these approaches to not give developers any information about how much performance gain is possible before the critical path changes.
Hollingsworth and Miller introduce two new metrics to approximate optimization potential: \emph{slack}, how much a procedure can be improved before the critical path changes; and \emph{logical zeroing}, the reduction in critical path length when a procedure is completely removed~\cite{hollingsworth1994slack}.
These metrics are similar to the optimization potential measured by a causal profiler, but can only be computed with a complete program activity graph.
Collection of a program activity graph is costly, and could introduce significant probe effect.


\paragraph{Bottleneck identification.}
Several approaches have used hardware performance counters to identify hardware-level performance bottlenecks~\cite{DBLP:journals/computer/MillerCCHIKKN95, DBLP:conf/sc/BurtscherKDMKB10, DBLP:conf/ispass/DiamondBMKKB11}.
Techniques based on binary instrumentation can identify cache and heap performance issues, contended locks, and other program hotspots~\cite{DBLP:conf/pldi/NethercoteS07, Bach:2010:APP:1749398.1749444, DBLP:conf/pldi/LukCMPKLWRH05}.
ParaShares and Harmony identify basic blocks that run during periods with little or no parallelism~\cite{DBLP:conf/europar/KambadurTK14, DBLP:conf/isca/KambadurTK12}.
Code identified by these tools is a good candidate for parallelization or classic serial optimizations.
Bottlenecks, a profile analysis tool, uses heuristics to identify bottlenecks using call-tree profiles~\cite{DBLP:conf/ecoop/AmmonsCGS04}.
Given call-tree profiles for different executions, Bottlenecks can pinpoint which procedures are responsible for the difference in performance.
The FreeLunch profiler and Visual Studio's contention profiler identify locks that are responsible for significant thread blocking time~\cite{DBLP:conf/oopsla/DavidTLM14, goldin2010thread}.
BIS uses similar techniques to identify highly contended critical sections on asymmetric multiprocessors, and automatically migrates performance-critical code to faster cores~\cite{DBLP:conf/asplos/JoaoSMP12}.
Bottle graphs present thread execution time and parallelism in a visual format that highlights program bottlenecks~\cite{DuBois:2013:BGV:2509136.2509529}.
Unlike causal profiling, these tools do not predict the performance impact of removing bottlenecks.
All these systems can only identify bottlenecks that arise from explicit thread communication, while causal profiling can measure parallel performance problems from any source, including cache coherence protocols, scheduling dependencies, and I/O.


\paragraph{Profiling for parallelization and scalability.}
Several systems have been developed to measure potential parallelism in serial programs~\cite{DBLP:conf/ppopp/PraunBC08, DBLP:conf/cgo/ZhangNJ09, DBLP:conf/pldi/GarciaJLT11}.
Like causal profiling, these systems identify code that will benefit from developer time.
Unlike causal profiling, these tools are not aimed at diagnosing performance issues in code that has already been parallelized.

Kulkarni, Pai, and Schuff present general metrics for available parallelism and scalability~\cite{DBLP:journals/sigmetrics/KulkarniPS10}.
The Cilkview scalability analyzer uses performance models for Cilk's constrained parallelism to estimate the performance effect of adding additional hardware threads~\cite{DBLP:conf/spaa/HeLL10}.
Causal profiling can detect performance problems that result from poor scaling on the current hardware platform.



\paragraph{Time attribution profilers.}
Time attribution profilers assign ``blame'' to concurrently executing
code based on what other threads are doing.  Quartz introduces the
notion of ``normalized processor time,'' which assigns high cost to
code that runs while a large fraction of other threads are
blocked~\cite{DBLP:conf/sigmetrics/AndersonL90}.  CPPROFJ extends this
approach to Java programs with aspects~\cite{DBLP:conf/kbse/Hall02}.
CPPROFJ uses finer categories for time: running, blocked for a
higher-priority thread, waiting on a monitor, and blocked on other
events.  Tallent and Mellor-Crummey extend this approach further to
support Cilk programs, with an added category for time spent managing
parallelism~\cite{DBLP:conf/ppopp/TallentM09}. The WAIT tool adds
fine-grained categorization to identify
bottlenecks in large-scale production Java
systems~\cite{DBLP:conf/oopsla/AltmanAFM10}.  Unlike causal profiling,
these profilers can only capture interference between threads that
directly affects their scheduler state.



\punt{
\paragraph{Tracing Profilers.}
Tracing profilers intercept interactions between threads or nodes in a
distributed system to construct a model of parallel performance. Monit
and the Berkeley UNIX Distributed Programs Monitor collect traces of
system-level events for later
analysis~\cite{DBLP:conf/sigmetrics/KerolaS87,DBLP:conf/icdcs/MillerMS85}.
\citet{DBLP:conf/sosp/AguileraMWRM03} use network-level tracing to
identify probable ``causal paths'' that may be responsible for high
latency in distributed systems of black boxes. AppInsight uses a
similar technique to identify sources of latency in mobile application
event handlers~\cite{ravindranath2012appinsight}.

Tracing is only practical in domains where thread interaction is
limited. Program threads may interact not only via synchronization
operations but also due to cache coherence protocols, race conditions,
scheduling dependencies, lock-free algorithms, and I/O, all of which
causal profiling implicitly accounts for.
}



\subsection{Performance Guidance and Experimentation}

Several systems have employed delays to extract information about program execution times.
Mytkowicz et al. use delays to validate the output of profilers on single-threaded Java programs~\cite{DBLP:conf/pldi/MytkowiczDHS10}.
Snelick, J\'aJ\'a et al. use delays to profile parallel programs~\cite{DBLP:journals/spe/SnelickJKL94}.
This approach measures the effect of slowdowns in combination, which requires a complete execution of the program for each of an exponential number of configurations.
Active Dependence Discovery (ADD) introduces performance perturbations to distributed systems and measures their impact on response time~\cite{DBLP:conf/im/BrownKK01}.
ADD requires a complete enumeration of system components, and requires developers to insert performance perturbations manually.
Gunawi, Agrawal et al. use delays to identify causal dependencies between events in the EMC Centera storage system to analyze Centera's protocols and policies~\cite{DBLP:conf/isca/GunawiAAAS05}.
Song and Lu use machine learning to identify performance anti-patterns in source code~\cite{DBLP:conf/oopsla/SongL14}.
Unlike causal profiling, these approaches do not predict the effect of potential optimizations.

%% file: conclusion.tex
\section{Conclusion}
\label{sec:conclusion}

Profilers are the primary tool in the programmer's toolbox for
identifying performance tuning opportunities. Previous profilers only
observe actual executions and correlate code
with execution time or performance counters. This information can be
of limited use because the amount of time
spent does not necessarily correspond to where programmers should
focus their optimization efforts. Past profilers are also limited to
reporting end-to-end execution time, an unimportant quantity for
servers and interactive applications whose key metrics of interest are
throughput and latency. Causal profiling is a new,
experiment-based approach that establishes causal relationships
between hypothetical optimizations and their effects. By virtually
speeding up lines of code, causal profiling identifies and quantifies the impact on
either throughput or latency of any degree of optimization to any line
of code. Our prototype causal profiler, \coz{}, is efficient,
accurate, and effective at guiding optimization efforts.

\punt{
\coz{} will be available for download at \url{anonymized.org}. 
}

%% file: main.bbl
\begin{thebibliography}{10}

\bibitem{DBLP:conf/oopsla/AltmanAFM10}
E.~R. Altman, M.~Arnold, S.~Fink, and N.~Mitchell.
\newblock Performance analysis of idle programs.
\newblock In {\em OOPSLA}, pages 739--753. ACM, 2010.

\bibitem{DBLP:conf/pldi/AmmonsBL97}
G.~Ammons, T.~Ball, and J.~R. Larus.
\newblock Exploiting hardware performance counters with flow and context
  sensitive profiling.
\newblock In {\em PLDI}, pages 85--96. ACM, 1997.

\bibitem{DBLP:conf/ecoop/AmmonsCGS04}
G.~Ammons, J.-D. Choi, M.~Gupta, and N.~Swamy.
\newblock Finding and removing performance bottlenecks in large systems.
\newblock In {\em ECOOP}, volume 3086 of {\em Lecture Notes in Computer
  Science}, pages 170--194. Springer, 2004.

\bibitem{DBLP:conf/sigmetrics/AndersonL90}
T.~E. Anderson and E.~D. Lazowska.
\newblock Quartz: A tool for tuning parallel program performance.
\newblock In {\em SIGMETRICS}, pages 115--125, 1990.

\bibitem{Bach:2010:APP:1749398.1749444}
M.~M. Bach, M.~Charney, R.~Cohn, E.~Demikhovsky, T.~Devor, K.~Hazelwood,
  A.~Jaleel, C.-K. Luk, G.~Lyons, H.~Patil, and A.~Tal.
\newblock Analyzing parallel programs with {Pin}.
\newblock {\em Computer}, 43(3):34--41, Mar. 2010.

\bibitem{DBLP:conf/micro/BallL96}
T.~Ball and J.~R. Larus.
\newblock Efficient path profiling.
\newblock In {\em MICRO}, pages 46--57, 1996.

\bibitem{DBLP:conf/im/BrownKK01}
A.~B. Brown, G.~Kar, and A.~Keller.
\newblock An active approach to characterizing dynamic dependencies for problem
  determination in a distributed environment.
\newblock In {\em Integrated Network Management}, pages 377--390. IEEE, 2001.

\bibitem{DBLP:conf/sc/BurtscherKDMKB10}
M.~Burtscher, B.-D. Kim, J.~R. Diamond, J.~D. McCalpin, L.~Koesterke, and J.~C.
  Browne.
\newblock {PerfExpert}: An easy-to-use performance diagnosis tool for {HPC}
  applications.
\newblock In {\em SC}, pages 1--11. IEEE, 2010.

\bibitem{DBLP:conf/cgo/ChoMHBM13}
H.~K. Cho, T.~Moseley, R.~E. Hank, D.~Bruening, and S.~A. Mahlke.
\newblock Instant profiling: Instrumentation sampling for profiling datacenter
  applications.
\newblock In {\em CGO}, pages 1--10. IEEE Computer Society, 2013.

\bibitem{stabilizer:asplos13}
C.~Curtsinger and E.~D. Berger.
\newblock {\sc Stabilizer}: Statistically sound performance evaluation.
\newblock In {\em ASPLOS}, New York, NY, USA, 2013. ACM.

\bibitem{DBLP:conf/oopsla/DavidTLM14}
F.~David, G.~Thomas, J.~Lawall, and G.~Muller.
\newblock Continuously measuring critical section pressure with the free-lunch
  profiler.
\newblock In {\em OOPSLA}, pages 291--307, 2014.

\bibitem{DBLP:conf/ispass/DiamondBMKKB11}
J.~R. Diamond, M.~Burtscher, J.~D. McCalpin, B.-D. Kim, S.~W. Keckler, and
  J.~C. Browne.
\newblock Evaluation and optimization of multicore performance bottlenecks in
  supercomputing applications.
\newblock In {\em ISPASS}, pages 32--43. IEEE Computer Society, 2011.

\bibitem{DuBois:2013:BGV:2509136.2509529}
K.~Du~Bois, J.~B. Sartor, S.~Eyerman, and L.~Eeckhout.
\newblock Bottle graphs: Visualizing scalability bottlenecks in multi-threaded
  applications.
\newblock In {\em OOPSLA}, pages 355--372, 2013.

\bibitem{gdb_manual}
Free Software Foundation.
\newblock {\em Debugging with {GDB}}, tenth edition.

\bibitem{DBLP:conf/pldi/GarciaJLT11}
S.~Garcia, D.~Jeon, C.~M. Louie, and M.~B. Taylor.
\newblock Kremlin: rethinking and rebooting gprof for the multicore age.
\newblock In {\em PLDI}, pages 458--469. ACM, 2011.

\bibitem{goldin2010thread}
M.~Goldin.
\newblock Thread performance: Resource contention concurrency profiling in
  visual studio 2010.
\newblock {\em MSDN magazine}, page~38, 2010.

\bibitem{DBLP:conf/sigplan/GrahamKM82}
S.~L. Graham, P.~B. Kessler, and M.~K. McKusick.
\newblock gprof: a call graph execution profiler.
\newblock In {\em SIGPLAN Symposium on Compiler Construction}, pages 120--126.
  ACM, 1982.

\bibitem{DBLP:conf/isca/GunawiAAAS05}
H.~S. Gunawi, N.~Agrawal, A.~C. Arpaci{-}Dusseau, R.~H. Arpaci{-}Dusseau, and
  J.~Schindler.
\newblock Deconstructing commodity storage clusters.
\newblock In {\em ISCA}, pages 60--71. {IEEE} Computer Society, 2005.

\bibitem{DBLP:conf/kbse/Hall02}
R.~J. Hall.
\newblock {CPPROFJ: Aspect-Capable Call Path Profiling of Multi-Threaded {Java}
  Applications}.
\newblock In {\em ASE}, pages 107--116. IEEE Computer Society, 2002.

\bibitem{DBLP:conf/spaa/HeLL10}
Y.~He, C.~E. Leiserson, and W.~M. Leiserson.
\newblock The {Cilkview} scalability analyzer.
\newblock In {\em SPAA}, pages 145--156. ACM, 2010.

\bibitem{DBLP:conf/pdp/HillJSV98}
J.~M.~D. Hill, S.~A. Jarvis, C.~J. Siniolakis, and V.~P. Vasilev.
\newblock Portable and architecture independent parallel performance tuning
  using a call-graph profiling tool.
\newblock In {\em PDP}, pages 286--294, 1998.

\bibitem{hollingsworth1994slack}
J.~K. Hollingsworth and B.~P. Miller.
\newblock Slack: a new performance metric for parallel programs.
\newblock {\em University of Maryland and University of Wisconsin-Madison,
  Tech. Rep}, 1994.

\bibitem{intel_parallel_studio}
Intel.
\newblock {\em {Intel VTune Amplifier 2015}}, 2014.

\bibitem{DBLP:conf/asplos/JoaoSMP12}
J.~A. Joao, M.~A. Suleman, O.~Mutlu, and Y.~N. Patt.
\newblock Bottleneck identification and scheduling in multithreaded
  applications.
\newblock In {\em ASPLOS}, pages 223--234. ACM, 2012.

\bibitem{DBLP:conf/isca/KambadurTK12}
M.~Kambadur, K.~Tang, and M.~A. Kim.
\newblock Harmony: Collection and analysis of parallel block vectors.
\newblock In {\em ISCA}, pages 452--463. {IEEE} Computer Society, 2012.

\bibitem{DBLP:conf/europar/KambadurTK14}
M.~Kambadur, K.~Tang, and M.~A. Kim.
\newblock Parashares: Finding the important basic blocks in multithreaded
  programs.
\newblock In {\em Euro-Par}, Lecture Notes in Computer Science, pages 75--86,
  2014.

\bibitem{perf_wiki}
kernel.org.
\newblock {\em {perf}: {Linux} profiling with performance counters}, 2014.

\bibitem{DBLP:journals/sigmetrics/KulkarniPS10}
M.~Kulkarni, V.~S. Pai, and D.~L. Schuff.
\newblock Towards architecture independent metrics for multicore performance
  analysis.
\newblock {\em SIGMETRICS Performance Evaluation Review}, 38(3):10--14, 2010.

\bibitem{oprofile}
J.~Levon and P.~Elie.
\newblock {oprofile}: A system profiler for {Linux}.
\newblock http://oprofile.sourceforge.net/, 2004.

\bibitem{little2011or}
J.~D. Little.
\newblock {OR FORUM: {Little}'s Law as Viewed on Its 50th Anniversary}.
\newblock {\em Operations Research}, 59(3):536--549, 2011.

\bibitem{DBLP:conf/pldi/LukCMPKLWRH05}
C.-K. Luk, R.~S. Cohn, R.~Muth, H.~Patil, A.~Klauser, P.~G. Lowney, S.~Wallace,
  V.~J. Reddi, and K.~M. Hazelwood.
\newblock Pin: building customized program analysis tools with dynamic
  instrumentation.
\newblock In {\em PLDI}, pages 190--200, 2005.

\bibitem{DBLP:journals/computer/MillerCCHIKKN95}
B.~P. Miller, M.~D. Callaghan, J.~M. Cargille, J.~K. Hollingsworth, R.~B.
  Irvin, K.~L. Karavanic, K.~Kunchithapadam, and T.~Newhall.
\newblock The paradyn parallel performance measurement tool.
\newblock {\em IEEE Computer}, 28(11):37--46, 1995.

\bibitem{DBLP:journals/tpds/MillerCHKLT90}
B.~P. Miller, M.~Clark, J.~K. Hollingsworth, S.~Kierstead, S.-S. Lim, and
  T.~Torzewski.
\newblock {IPS-2}: The second generation of a parallel program measurement
  system.
\newblock {\em IEEE Transactions on Parallel Distributed Systems},
  1(2):206--217, 1990.

\bibitem{DBLP:conf/icdcs/MillerY87}
B.~P. Miller and C.-Q. Yang.
\newblock {IPS}: An interactive and automatic performance measurement tool for
  parallel and distributed programs.
\newblock In {\em ICDCS}, pages 482--489, 1987.

\bibitem{DBLP:conf/pldi/MytkowiczDHS10}
T.~Mytkowicz, A.~Diwan, M.~Hauswirth, and P.~F. Sweeney.
\newblock Evaluating the accuracy of {Java} profilers.
\newblock In {\em PLDI}, pages 187--197. ACM, 2010.

\bibitem{DBLP:conf/pldi/NethercoteS07}
N.~Nethercote and J.~Seward.
\newblock Valgrind: a framework for heavyweight dynamic binary instrumentation.
\newblock In {\em PLDI}, pages 89--100. ACM, 2007.

\bibitem{DBLP:conf/ipps/OyamaTY00}
Y.~Oyama, K.~Taura, and A.~Yonezawa.
\newblock Online computation of critical paths for multithreaded languages.
\newblock In {\em IPDPS Workshops}, volume 1800 of {\em Lecture Notes in
  Computer Science}, pages 301--313. Springer, 2000.

\bibitem{DBLP:journals/spe/SnelickJKL94}
R.~Snelick, J.~J{\'a}J{\'a}, R.~Kacker, and G.~Lyon.
\newblock Synthetic-perturbation techniques for screening shared memory
  programs.
\newblock {\em Software Practice \& Experience}, 24(8):679--701, 1994.

\bibitem{DBLP:conf/oopsla/SongL14}
L.~Song and S.~Lu.
\newblock Statistical debugging for real-world performance problems.
\newblock In {\em OOPSLA}, pages 561--578, 2014.

\bibitem{DBLP:conf/sc/SzebenyiWW09}
Z.~Szebenyi, F.~Wolf, and B.~J.~N. Wylie.
\newblock Space-efficient time-series call-path profiling of parallel
  applications.
\newblock In {\em SC}. ACM, 2009.

\bibitem{DBLP:conf/ppopp/TallentM09}
N.~R. Tallent and J.~M. Mellor-Crummey.
\newblock Effective performance measurement and analysis of multithreaded
  applications.
\newblock In {\em PPOPP}, pages 229--240. ACM, 2009.

\bibitem{thompson1975unix}
K.~Thompson and D.~M. Ritchie.
\newblock {\em UNIX Programmer's Manual}.
\newblock Bell Telephone Laboratories, 1975.

\bibitem{DBLP:conf/ppopp/PraunBC08}
C.~von Praun, R.~Bordawekar, and C.~Cascaval.
\newblock Modeling optimistic concurrency using quantitative dependence
  analysis.
\newblock In {\em PPoPP}, pages 185--196. ACM, 2008.

\bibitem{DBLP:journals/tse/YangM89}
C.-Q. Yang and B.~P. Miller.
\newblock Performance measurement for parallel and distributed programs: A
  structured and automatic approach.
\newblock {\em IEEE Transactions on Software Engineering}, 15(12):1615--1629,
  1989.

\bibitem{DBLP:conf/cgo/ZhangNJ09}
X.~Zhang, A.~Navabi, and S.~Jagannathan.
\newblock Alchemist: A transparent dependence distance profiling
  infrastructure.
\newblock In {\em CGO}, pages 47--58. IEEE Computer Society, 2009.

\end{thebibliography}
